\documentclass[twocolumn]{aastex631}

\usepackage{CJK}
\usepackage{soul}
\usepackage{mathtools}
\usepackage{makecell}
\usepackage{amsmath}
\usepackage{multirow}
\usepackage{bbold}

\begin{document}

\title{Galaxy Spin Classification I: Z-wise vs S-wise Spirals With Chirality Equivariant Residual Network}

\correspondingauthor{He Jia}
\email{hejia@princeton.edu}

\author[0000-0002-9958-7758]{He Jia \begin{CJK*}{UTF8}{gbsn}(贾赫)\end{CJK*}}
\affiliation{Department of Astrophysical Sciences, Princeton University, Princeton, NJ 08544, USA}

\author[0000-0002-8202-8642]{Hong-Ming Zhu \begin{CJK*}{UTF8}{gbsn}(朱弘明)\end{CJK*}}
\affiliation{Canadian Institute for Theoretical Astrophysics, 60 St. George Street, Toronto, Ontario M5S 3H8, Canada}

\author[0000-0003-2155-9578]{Ue-Li Pen \begin{CJK*}{UTF8}{bsmi}(彭威禮)\end{CJK*}}
\affiliation{Institute of Astronomy and Astrophysics, Academia Sinica, Astronomy-Mathematics Building, No. 1, Sec. 4, Roosevelt Road, Taipei 10617, Taiwan}
\affiliation{Canadian Institute for Theoretical Astrophysics, 60 St. George Street, Toronto, Ontario M5S 3H8, Canada}
\affiliation{Canadian Institute for Advanced Research, 661 University Ave, Toronto, Ontario M5G 1M1, Canada}
\affiliation{Dunlap Institute for Astronomy and Astrophysics, University of Toronto, 50 St. George Street, Toronto, Ontario M5S 3H4, Canada}
\affiliation{Perimeter Institute of Theoretical Physics, 31 Caroline Street North, Waterloo, Ontario N2L 2Y5, Canada}



\begin{abstract}

The angular momentum of galaxies (galaxy \textit{spin}) contains rich information about the initial condition of the Universe, yet it is challenging to efficiently measure the spin direction for the tremendous amount of galaxies that are being mapped by the ongoing and forthcoming cosmological surveys.
We present a machine learning based classifier for the Z-wise vs S-wise spirals, which can help to break the degeneracy in the galaxy spin direction measurement.
The proposed Chirality Equivariant Residual Network (CE-ResNet) is manifestly equivariant under a reflection of the input image, which guarantees that there is no inherent asymmetry between the Z-wise and S-wise probability estimators.
We train the model with Sloan Digital Sky Survey (SDSS) images, with the training labels given by the Galaxy Zoo 1 (GZ1) project.
A combination of data augmentation tricks are used during the training, making the model more robust to be applied to other surveys.
We find a $\sim\!30\%$ increase of both types of spirals when Dark Energy Spectroscopic Instrument (DESI) images are used for classification, due to the better imaging quality of DESI.
We verify that the $\sim\!7\sigma$ difference between the numbers of Z-wise and S-wise spirals is due to human bias, since the discrepancy drops to $<\!1.8\sigma$ with our CE-ResNet classification results.
We discuss the potential systematics that are relevant to the future cosmological applications.


\end{abstract}



\section{Introduction}

\label{sec:intro}




In tidal-torque theory, the angular momentum of galaxies (galaxy \textit{spin}) is generated by the tidal torque due to the misalignment between the protohalo inertia tensor and the local gravitational tidal shear \citep{1969ApJ...155..393P,1970Ap......6..320D,1984ApJ...286...38W}.
Cosmological simulations have confirmed that the direction of dark matter halo spin is well described by the tidal-torque theory \citep{2002MNRAS.332..325P},
and that disk galaxies generally follow dark matter and gain similar spin directions as their host halos \citep{2015ApJ...812...29T,2019MNRAS.488.4801J}.
This makes galaxy spin a promising cosmological probe of various parameters including the initial condition of the Universe, primordial chirality and the neutrino mass \citep[see e.g.][for recent investigations]{2000ApJ...532L...5L,2001ApJ...555..106L,2019PhRvD..99l3532Y,2020PhRvL.124j1302Y,2021NatAs...5..283M,2022PhRvD.105h3512M,2022PhRvD.105h3504M}.

Recently, \citet{2021NatAs...5..283M} find a correlation between the observed galaxy spins and the initial density field of the Universe.
The same galaxy catalog has also been used to search for primordial chirality violations \citep{2022PhRvD.105h3512M}.
However, the signal to noise ratio is limited by the number of galaxies ($\sim15000$) with their spin directions observed.
In order to optimally extract the information from galaxy spin data to constrain the evolution of our universe,
it is necessary to develop new methods to measure the direction of galaxy spin accurately and efficiently.

Assuming that spiral galaxies are well approximated by circular disks, their three-dimensional spin directions can be determined through the position angles and axis ratios which are readily available from photometric observations, up to a fourfold degeneracy \citep[see Figure 2 of][for a visual illustration]{2021NatAs...5..283M}.
\citet{iye2019spin} visually inspect 842 spiral galaxies and confirm that (1) all the spirals are \textit{trailing}, i.e. Z-wise spirals rotate clockwisely, and (2) the dark, dust-lane-dominant side of the minor axis is closer to us.
Therefore, one can break the fourfold degeneracy if one can determine (1) whether the galaxy is a Z-wise or S-wise spiral, and (2) which side of the minor axis is darker and redder.\footnote{Alternatively, one can use the information regarding which side of the major axis is approaching us \citep[][]{1995ApJ...445...46H,2000ApJ...543L.107P,2021NatAs...5..283M}. This may be less ambiguous than visually deciding the dark side of the minor axis, but requires spectroscopic data and thus cannot be directly obtained from photometric surveys.}
In this paper, we focus on the classification of Z-wise vs S-wise spirals, and leave dark side vs bright side classification for future research.

Galaxy Zoo 1 (GZ1) is a citizen science project which classifies about $9\times10^5$ galaxies by members of the public \citep[][]{lintott2008galaxy,lintott2011galaxy}.
\footnote{\url{http://zoo1.galaxyzoo.org/}}
It provides the information of clockwise or anticlockwise (Z-wise or S-wise spiral pattern) for galaxies from the Sloan Digital Sky Survey \citep[SDSS,][]{abazajian2009seventh,ahumada202016th} data, which leads to a $\sim3\sigma$ detection of the correlation between the galaxy spin field and cosmological initial conditions \citep[][]{2021NatAs...5..283M} and preliminary results for primordial chirality violations \citep[][]{2022PhRvD.105h3512M}.
However, the ongoing and forthcoming cosmological surveys, such as the Dark Energy Spectroscopic Instrument \citep[DESI,][]{aghamousa2016desi1,aghamousa2016desi2}, will map tens of more galaxies than those classified in GZ1, which are prohibitive to be again classified by human.
Machine learning (ML) based classification methods are required to efficiently identify the morphological properties of galaxies.


Deep Convolutional Neural Networks (CNN) have led to a series of breakthroughs in computer vision during the past ten years \citep{lecun1989backpropagation,krizhevsky2012imagenet,he2016deep,tan2019efficient}.
They are now regarded as the state-of-the-art image classification method and are widely used in general astrophysics applications  \citep{2010MNRAS.406..342B,2011A&A...525A.157H,2015MNRAS.450.1441D,he2019learning,2020arXiv201213083A,lin2020feature}.
In this paper, we present Chirality Equivariant Residual Network (CE-ResNet), a machine learning based classifier for the Z-wise vs S-wise spirals.
The remainder of this paper is organized as follows: we introduce the datasets in Section \ref{sec:dataset} and the architecture of our model in Section \ref{sec:arch}.
We present the training and classification results of our model in Section \ref{sec:train} with SDSS images, and in Section \ref{sec:desi-image} with DESI images.
The known asymmetry between Z-Spirals and S-Spirals in GZ1 is discussed in Section \ref{sec:violation}.
We conclude this paper in Section \ref{sec:discuss}.
The source code of our CE-ResNet model \footnote{\url{https://github.com/h3jia/galaxy_spin_classifier/}} and the classification catalogs \footnote{\dataset[DOI: 10.5281/zenodo.7170929]{https://doi.org/10.5281/zenodo.7170929}} are publicly available.

\section{Galaxy Spin Datasets}

\label{sec:dataset}

\begin{deluxetable*}{cccccccc}

\label{table:pzps}
\tablecaption{The numbers of Z-wise and S-wise spirals in the different classification catalogs.
A galaxy is classified as a Z(S)-wise spiral if its $p_{\rm z}$($p_{\rm s}$) is larger than the cutoff value $p_{\rm cut}$.
See the corresponding sections for the details of each combination.
We also list the significance values for chirality violation based on the numbers of Z-wise and S-wise spirals; see Section \ref{sec:violation} for more details.}
\def\arraystretch{1.1}

\tablehead{Catalog & Image Source & Classifier & $p_{\rm cut}$ & Z-wise & S-wise & $\sigma$-value & See}

\startdata
\multirow{5}{*}{Reduced GZ1} & \multirow{5}{*}{SDSS} & \multirow{5}{*}{GZ1 Humans} & 0.5 & 35068 & 37022 & 7.278 & \multirow{5}{*}{Section \ref{sec:dataset}} \\
& & & 0.6 & 27753 & 29535 & 7.445 & \\
& & & 0.7 & 22064 & 23488 & 6.672 & \\
& & & 0.8 & 16674 & 17921 & 6.704 & \\
& & & 0.9 & 10269 & 11283 & 6.907 & \\
\hline
\multirow{5}{*}{Reduced GZ1} & \multirow{5}{*}{SDSS} & \multirow{5}{*}{CE-ResNet} & 0.5 & 35327 & 35667 & 1.276 & \multirow{5}{*}{Section \ref{sec:train}} \\
& & & 0.6 & 27437 & 27737 & 1.277 & \\
& & & 0.7 & 21036 & 21283 & 1.201 & \\
& & & 0.8 & 15433 & 15542 & 0.619 & \\
& & & 0.9 & 9218 & 9442 & 1.640 & \\
\hline
\multirow{5}{*}{Reduced GZ1} & \multirow{5}{*}{DESI} & \multirow{5}{*}{CE-ResNet} & 0.5 & 44362 & 45432 & 3.571 & \multirow{5}{*}{Section \ref{sec:desi-image}} \\
& & & 0.6 & 36892 & 37619 & 2.663 & \\
& & & 0.7 & 29826 & 30217 & 1.596 & \\
& & & 0.8 & 22525 & 22804 & 1.310 & \\
& & & 0.9 & 13583 & 13775 & 1.161 & \\
\hline
\multirow{5}{*}{Preliminary DESI} & \multirow{5}{*}{DESI} & \multirow{5}{*}{CE-ResNet} & 0.5 & 55243 & 55526 & 0.850 & \multirow{5}{*}{Section \ref{sec:pre-desi}} \\
& & & 0.6 & 41977 & 42269 & 1.006 & \\
& & & 0.7 & 31508 & 31693 & 0.736 & \\
& & & 0.8 & 22012 & 22143 & 0.623 & \\
& & & 0.9 & 11649 & 11919 & 1.759 & \\
\enddata

\end{deluxetable*}

In the GZ1 project, public volunteers are asked to classify the SDSS galaxy images into six categories: ellipticals, Z-wise spirals, S-wise spirals, edge-on spirals, star / don't know, and mergers.
The catalog includes the vote counts of the six morphological types for 667,944 galaxies with SDSS spectra data available, and for 225,268 galaxies with no spectra data available.
The empirical probability that a galaxy belongs to each category can be estimated by the fraction of votes, while only the probabilities of Z-wise and S-wise spirals, $p_{\rm z}$ and $p_{\rm s}$, are relevant for this paper.
\footnote{Note that $p_{\rm z/s}$ throughout this paper is the probability that a random volunteer in the GZ1 project will decide that the galaxy is a Z(S)-Spiral, which is not exactly the same as the probability that the galaxy is actually a Z(S)-Spiral.
However, these two types of probabilities should be positively correlated, and the galaxies with $p_{\rm z/s}\to1$ are indeed very likely to be real Z(S)-Spirals (see the sample images in Appendix \ref{sec:sample}).
Therefore, the empirical probability estimated by the vote fractions is still a useful quantity for the classification of galaxy morphology.}
For simplicity, we effectively treat all the mergers as Non-Spirals, based on their vote fractions, despite that some mergers may indeed contain Z(S)-Spirals.
\footnote{If one object has a large $p_{\rm merger}$, its $p_{\rm z/s}$ should be small, since the vote fractions of the six categories add up to 1.}
We refer to the empirical probabilities from the GZ1 catalog as the ``true" probability, in contrast to the ``predicted" probability given by the classifiers.
To keep the number of different kinds of galaxies roughly balanced, we downsample the galaxies with $p_{\rm m}\equiv\max\,(p_{\rm z},p_{\rm s}) \in [0,0.1]$ by a factor of 20 (i.e. only keep 1 of 20 such galaxies), the galaxies with $p_{\rm m} \in (0.1,0.2]$ by a factor of 5, and the galaxies with $p_{\rm m} \in (0.2,0.3]$ by a factor of 2.
We then query the SDSS DR16 SQL database with the RA and Dec of each galaxy in the GZ1 catalog, and apply the following cuts to remove the galaxies that are unlikely to be clear enough to identify their morphology:
\begin{enumerate}
    \item There should be exactly one PhotoObj within $1"$ of the location in the GZ1 catalog.
    \item The error of $r$ band magnitude should be in $(0,1)$.
    \item The $r$ band half light radius $r_{50}$ should be larger than $1"$.
    \item The $r$ band relative error of radius $\sigma_{r_{50}}/r_{50}$ should be in $(0",0.25")$.
\end{enumerate}
We find 173,097 galaxies that meet all the criteria (dubbed ``Reduced GZ1" catalog henceforth), and use $70\%$ of them for training, $15\%$ for validation, and $15\%$ for testing.
We obtain the jpeg images of these galaxies from both SDSS and DESI surveys using the Legacy Surveys Sky Viewer tool \citep{dey2019overview}.
\footnote{\url{https://www.legacysurvey.org/}}
The SDSS DR16 images are generated from the \textit{gri} bands, while the DESI DR9 images are generated from the \textit{grz} bands.
The numbers of $p_{\rm z}$ and $p_{\rm s}$ galaxies with different choices of cutoff values $p_{\rm cut}$ for the vote fraction are listed in Table \ref{table:pzps}.
Sample images for galaxies with different GZ1 morphology classification probabilities can be found in Figure \ref{fig:sam-gz-sdss}.

To assess whether our model can be robustly applied to surveys other than SDSS (with which the model is trained), we also collect all the galaxies in the DESI Legacy Survey Sweep Catalogs that are larger than $1"$ in half light radius and have spectroscopic redshifts available, leading to the ``Preliminary DESI" catalog of 1,953,246 galaxies.
For our model, the image field-of-view (FOV) is chosen as a multiple of the galaxy size, which however may deviate significantly between different survey measurements (see Figure \ref{fig:desi-spec}).
This requires that our classification model should be insensitive to the image FOV, which will be justified later in this paper (see Figure \ref{fig:stable} and the discussions therein).

\section{Network Architecture}

\label{sec:arch}

\begin{deluxetable}{ccc}

\label{table:resnet50}
\tablecaption{The structure of our Chirality Equivariant Residual Network (CE-ResNet).
It is based on the original ResNet-50 network in \citet{he2016deep}, but has four additional fully connected layers.
The network predicts the scores of Z-Spirals and Non-Spirals from the original images, and the scores of S-Spirals and Non-Spirals from the flipped images, which guarantees that it is equivariant under a parity inversion.
See Figure \ref{fig:arch} for a demonstration of the model workflow.}
\def\arraystretch{1.1}

\tablehead{
\colhead{layer name} & \colhead{output size} & \colhead{54-layer CE-ResNet}
}

\startdata
conv1 & $80\times80\times64$ & $7\times7$, 64, stride 2 \\[0.05cm]
\hline \\[-0.35cm]
conv2\_x & $40\times40\times256$ & \makecell{$3\times3$ max pool, stride 2 \vspace{0.1cm} \\ $\left[
\begin{tabular}{c}
    1$\times$1, 64 \\[0.1cm]
    3$\times$3, 64 \\[0.1cm]
    1$\times$1, 256 \\
\end{tabular}
\right] \times 3$} \\[0.9cm]
\hline \\[-0.3cm]
conv3\_x & $20\times20\times512$ & $\left[
\begin{tabular}{c}
    1$\times$1, 128 \\
    3$\times$3, 128 \\
    1$\times$1, 512 \\
\end{tabular}
\right] \times 4$ \\[0.7cm]
\hline \\[-0.3cm]
conv4\_x & $10\times10\times1024$ & $\left[
\begin{tabular}{c}
    1$\times$1, 256 \\
    3$\times$3, 256 \\
    1$\times$1, 1024 \\
\end{tabular}
\right] \times 6$ \\[0.7cm]
\hline \\[-0.3cm]
conv5\_x & $5\times5\times2048$ & $\left[
\begin{tabular}{c}
    1$\times$1, 512 \\
    3$\times$3, 512 \\
    1$\times$1, 2048 \\
\end{tabular}
\right] \times 3$ \\[0.7cm]
\hline \\[-0.35cm]
fc & $2$ & \makecell{average pool, output $1\times1\times2048$ \vspace{0.1cm} \\ 512-512-64-64-2 fully connected} \\[0.4cm]
\enddata

\end{deluxetable}

\begin{figure}
    \epsscale{1.045}
    \plotone{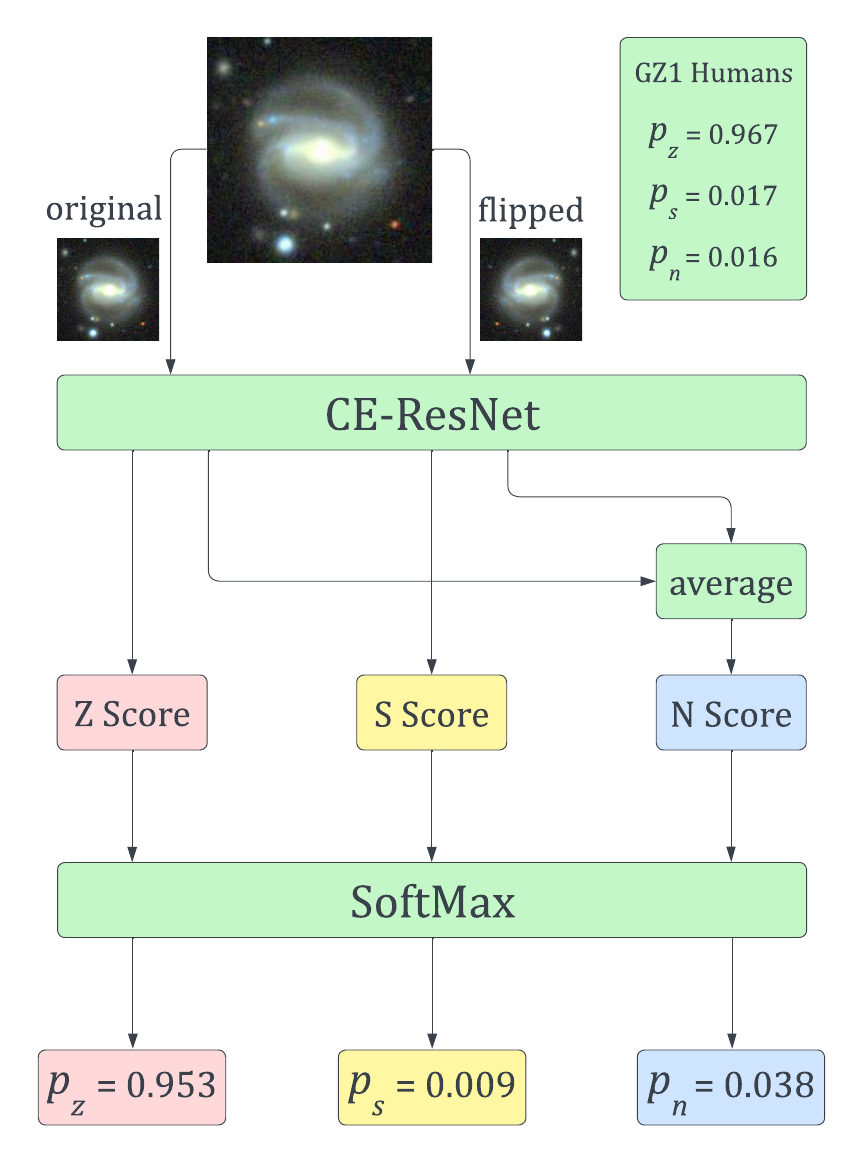}
    \caption{The workflow of the CE-ResNet model.
    Each galaxy image is fed into the network twice: given the original image the network outputs the scores for Z-Spirals and Non-Spirals, while given the flipped image the network outputs the scores for S-Spirals and Non-Spirals.
    We then average the two output scores for Non-Spirals and apply a softmax function to get the probabilities for the three categories.}
    \label{fig:arch}
\end{figure}

Empirically, deeper neural networks are more expressive than their shallower analogs, at the cost of being more difficult to train \citep{mhaskar2017when,mehta2019high}:
as the depth of the network increases, its accuracy may get saturated and then eventually decreases rapidly, which is known as the \textit{degradation} problem \citep{he2015convolutional,srivastava2015training}.
This issue can be mitigated by deep Residual Networks \citep[ResNet,][]{he2016deep}, which allows the construction of extremely deep convolution networks and is still considered as the state-of-the-art method for computer vision tasks.
The key insight of ResNet is that, in principle deeper networks should perform at least as well as their shallower analogs, since if the subsequent layers are all identity, a deep network becomes equivalent to a shallow network.
The degradation problem then implies that it is not easy for neural networks to approximate the \textit{identity} function with nonlinear layers.
Therefore, instead of fitting the target function $\mathcal{H}(x)$ directly, it is advantageous to have each nonlinear layer fit $\mathcal{F}(x) \coloneqq \mathcal{H}(x)-x$, which can be implemented by a simple \textit{shortcut} connection \citep[see e.g. Figure 2 of][]{he2016deep}.
Then, if the linear weights and biases are all zero, we have $\mathcal{F}(x)\equiv0$ and $\mathcal{H}(x)$ is identity.
This leads to an explicit solution for deeper networks such that they should perform no worse than shallower networks.

\begin{figure*}
    \plotone{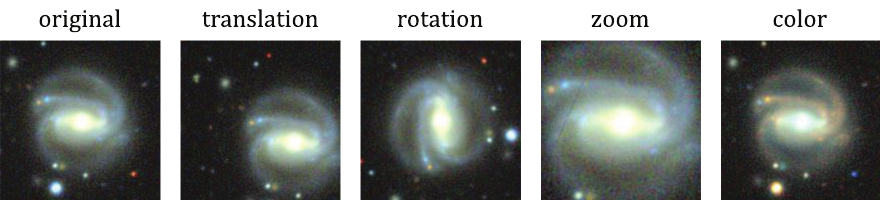}
    \caption{The four types of data augmentation used during the training of the network, which increases the effective data size and mitigates overfitting to the irrelevant information in the galaxy images.}
    \label{fig:data-aug}
\end{figure*}

\begin{figure*}
    \epsscale{0.64}
    \plotone{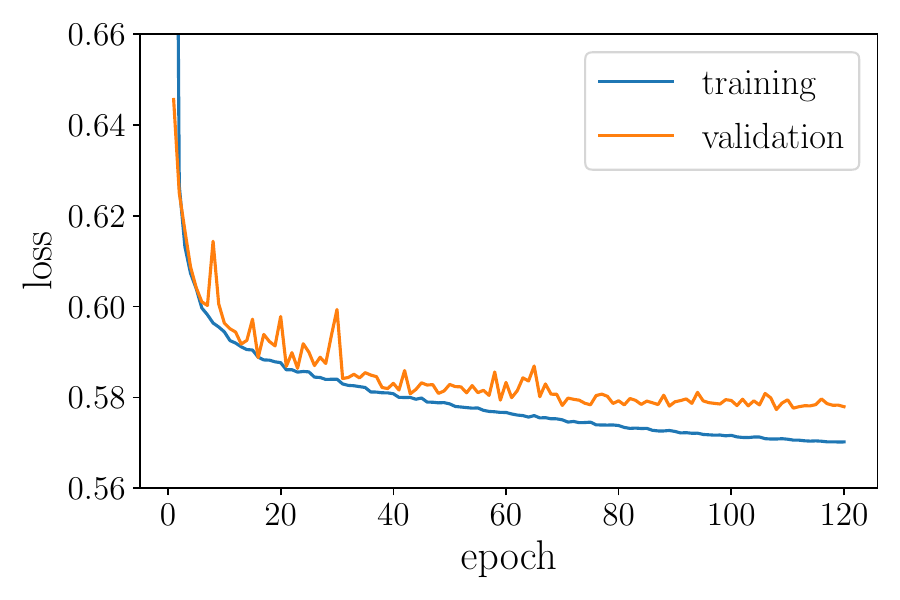}
    \caption{The training loss and validation loss of out network, evaluated on the Reduced GZ1 dataset with SDSS images.
    We terminate the training after 120 epochs as the validation loss no longer decreases.}
    \label{fig:loss-epoch}
\end{figure*}

We implement our Chirality Equivariant Residual Network (CE-ResNet) in \texttt{pytorch} \citep{pytorch}, based on the ResNet-50 model in \citet{he2016deep}.
The network structure is shown in Table \ref{table:resnet50}.
We use ReLU activation function for the convolution layers, and tanh activation function for the fully connected layers.
Our model is the same as the original ResNet-50 model, except for the following changes.
(1) The input image size is now $3 \times 160 \times 160$, and the output size for each layer is changed accordingly.
(2) We add four additional fully connected layers to improve its expressivity.
(3) Each galaxy image is fed into the neural network twice: the same network predicts the scores for Z-Spirals and Non-Spirals given the original image, and the scores for S-Spirals and Non-Spirals given the flipped image.
We then average the two estimates of the Non-Spiral score, and apply a softmax function to get the probabilities for the three categories (see Figure \ref{fig:arch}).
This guarantees that the network is purely parity-even, which is crucial for the cosmological galaxy spin analysis.
In practice we flip the images about the vertical axis (i.e. left to right), as illustrated in Figure \ref{fig:arch}, while we note that the direction of flip should not affect the performance of our model, which is insensitive to the orientation of input image (see Figure \ref{fig:stable}).

Note that we directly use the network to predict the scores and probabilities; we do not select a cutoff value and divide the galaxies into discrete categories of Z-Spirals, S-Spirals and Non-Spirals before training.
Unlike standard ML datasets such as MNIST \citep{lecun1998gradient}, where it is obvious whether one image is a handwritten number 8 or not, many galaxy images are not clear enough so that one can assert they belong to a certain morphology category.
Although galaxies with larger $p_{\rm z/s}$ are more likely to be real Z/S-spirals, there is no simple cutoff value such that all the galaxies with $p_{\rm z/s}$ above it are Z/S-spirals while all the galaxies below it are not: as shown in Figure \ref{fig:sam-gz-sdss}, the images just change continuously with respect to $p_{\rm z/s}$.
The value of $p_{\rm z/s}$ in the GZ1 catalog is indeed a stochastic variable: assuming that the vote on one single galaxy from each person follows some i.i.d. distribution based on its morphology and image quality, the fraction of total votes for each category does asymptotically converge with infinite vote size, but will always have finite noise when the vote size is limited.
If one pre-divides the galaxies into discrete categories, there are always mislabelled galaxies around the cutoff probability, which will be confusing to the classifier.
Therefore, we stick with estimating the scores directly, and leave the interpretation of the output probabilities to the user.

\section{Training and Results}

\label{sec:train}

\begin{figure*}
    \epsscale{1.15}
    \plotone{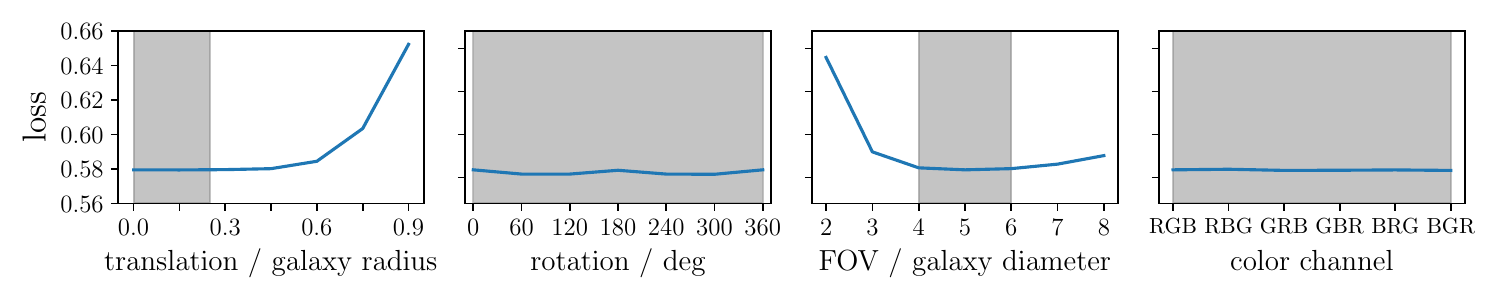}
    \caption{We check the stability of our model, under a translation, rotation, zoom and permutation of color channels.
    The grey bands indicate the data augmentation used in the training of our network.
    Our model is completely insensitive to the rotation and color channel permutation of input images.
    Its performance does degrade when the images are translated by more than 0.7 galaxy radii or when the FOV is smaller than 3 galaxy diameters, probably because some useful information is cut out from the images.
    On the other hand, using a larger FOV has a smaller impact on the classification results.}
    \label{fig:stable}
\end{figure*}

\begin{figure*}
    \epsscale{1.1}
    \plotone{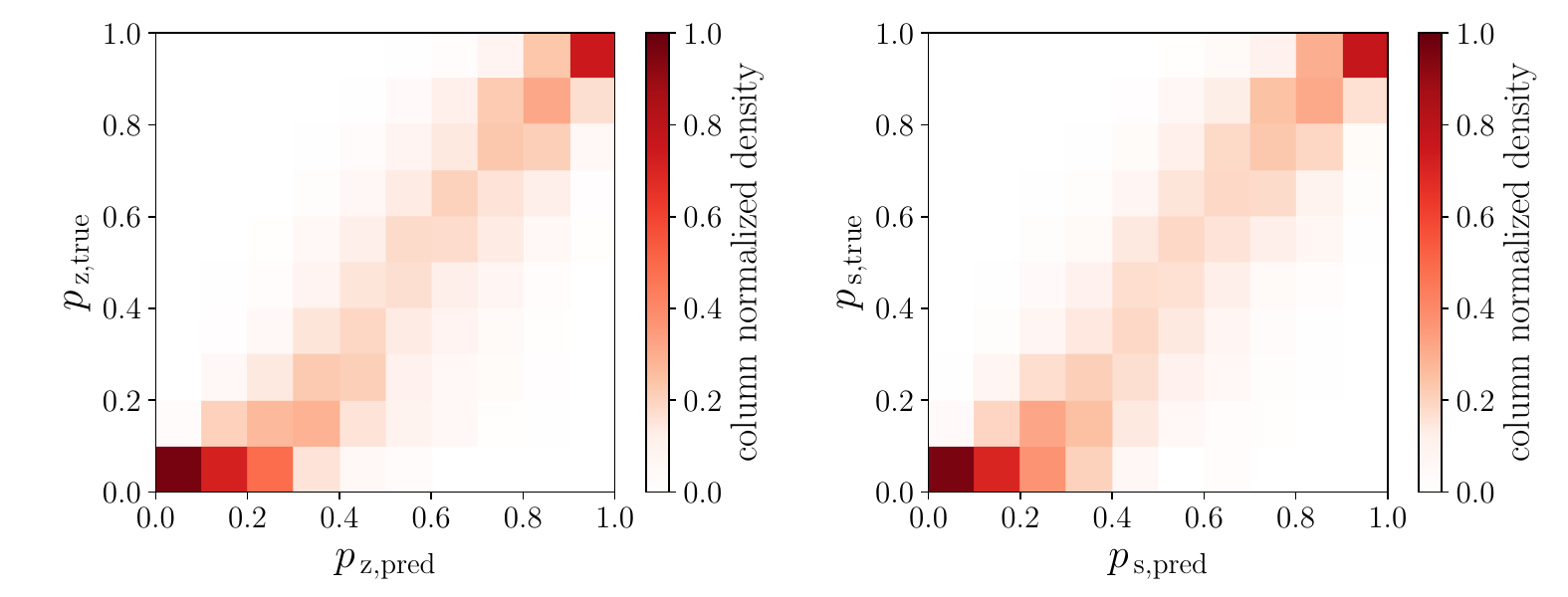}
    \caption{2-dim histograms for the true and predicted $p_{\rm z/s}$, with the density normalized by each column.
    The heat map is close to the ideal diagonal case, with the dispersion likely dominated by the Poisson noise in the training data.}
    \label{fig:heat}
\end{figure*}

The input of our model are 3-channel RGB images of shape $3\times160\times160$, with the target galaxy centered and the image FOV equal to five times the galaxy half light diameter.
Since the chirality of a spiral galaxy should have no dependence on its location, orientation, size and color, we apply the following data augmentation during training, which are illustrated in Figure \ref{fig:data-aug}:
\begin{enumerate}
    \item The relative location of the galaxy is moved by up to 25\% of its half light radius along both directions, with the exact translation randomly sampled from the uniform distribution $U(-0.25\,r_{50},\,0.25\,r_{50})$.
    \item The galaxy is rotated by a random angle, sampled from the uniform distribution $U(0^{\circ},\,360^{\circ})$.
    \item The exact FOV of training images is sampled from a uniform distribution between four times and six times the galaxy half light diameter.
    \item We apply a random permutation to the three color channels, to mitigate the potential overfitting to the correlation between galaxy color and morphology \citep[e.g.][]{bamford2009galaxy}. This will also improve the model generalizability to images from other surveys, with possibly different filters, instrumental responses and color scales.
\end{enumerate}
Here we stick with 3-channel images and do not average the different bands or use only the band with the best signal-to-noise ratio: although the color of galaxy should contain no chirality information, it is possible that e.g. the spiral arms where more new stars are forming look bluer than the region between the arms, so that the spiral structure is less clear in single-channel images compared with the full 3-channel images.


\begin{figure*}
    \epsscale{1.08}
    \plotone{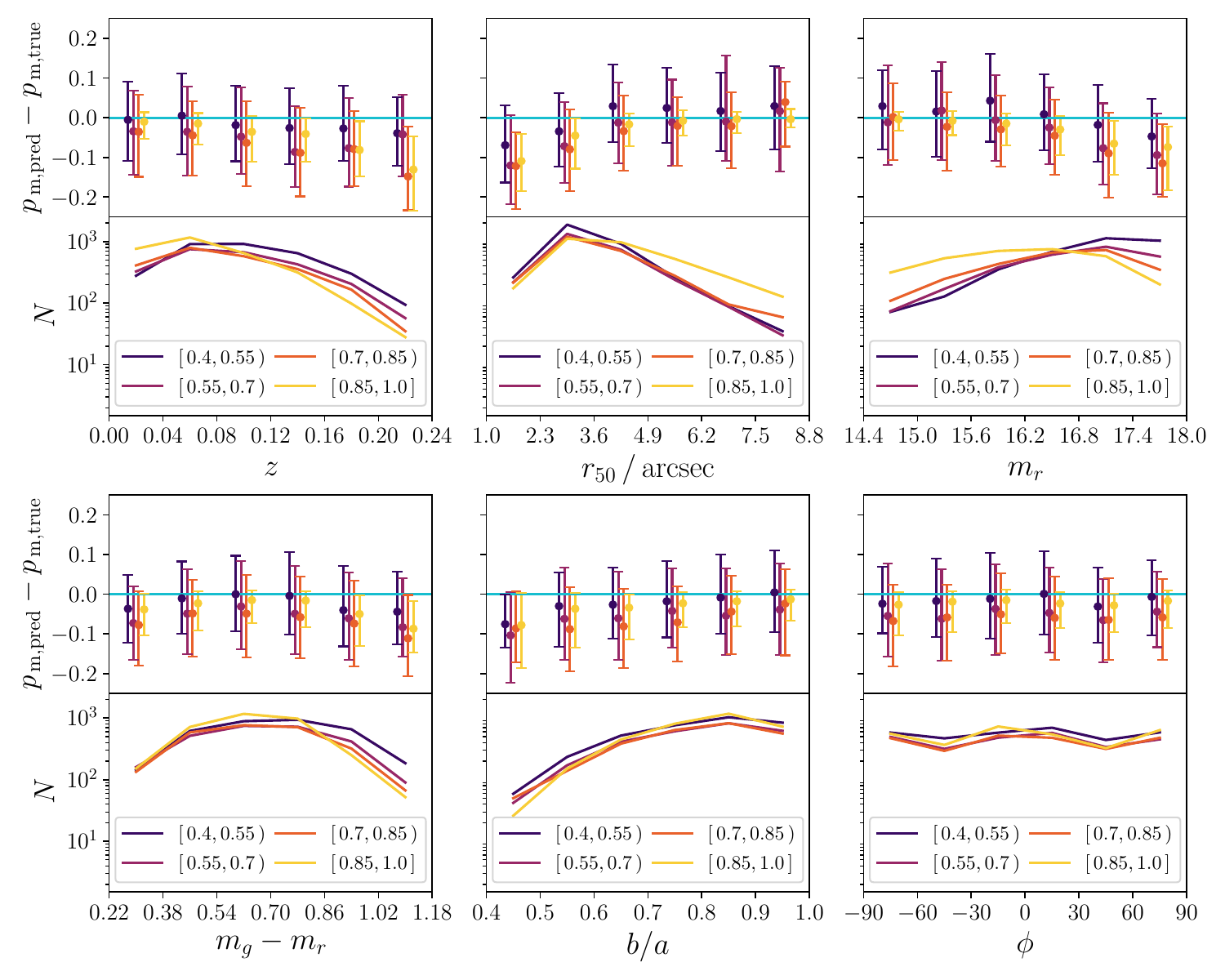}
    \caption{The quantiles of $p_{\rm m} \equiv \max(p_{\rm z},p_{\rm s})$ prediction errors for galaxies within different bins of redshift $z$, half light radius $r_{50}$, $r$ band magnitude $m_r$, color $m_g-m_r$, aspect ratio $b/a$, and orientation $\phi$, with the corresponding number counts of galaxies in each bin.
    The legend shows the four bins for $p_{\rm m}$.
    We note that for the same $p_{\rm m}$ bin, the $p_{\rm m}$ prediction is roughly unbiased with the variance independent of the six parameters, except that the model slightly underestimate $p_{\rm m}$ for the galaxies that are distant visually, visually small, dim, and have a small aspect ratio.
    Since our CE-ResNet model is manifestly parity-even, we find similar results if we plot $p_{\rm z}$ and $p_{\rm s}$ separately instead of $p_{\rm m}$.}
    \label{fig:sys}
\end{figure*}

We train our CE-ResNet model for 120 epochs using the cross entropy loss and the AdamW optimizer \citep{loshchilov2017decoupled}, which takes about one day on one NVIDIA V100 GPU.
We set the weight decay coefficient to 1 and use an initial learning rate of 0.0001, and after every 5 epochs we reduce the learning rate by a factor of $15\%$.
The training and validation losses are shown in Figure \ref{fig:loss-epoch}.
Sample images of different $p_{\rm z/s,pred}$ are shown in Figure \ref{fig:sam-cnn-sdss}, which are similar to those from the original GZ1 classification (Figure \ref{fig:sam-gz-sdss}).
The total number of spiral galaxies is also close to the original human classification (Table \ref{table:pzps}).

We test the robustness of our model under the data augmentation transforms in Figure \ref{fig:stable}, with only one type of transform activated each time.
Note that in the first panel, we apply a stochastic transform similar to the one used during training, with the maximum translation indicated by the horizontal axis.
For the other three panels, the same deterministic transform is applied to all the images in each case.
The loss function is almost constant within the extent of training data augmentation indicated by the grey bands, meaning that our model is indeed insensitive to these transforms.
Beyond the grey bands, the loss function does increase when a large translation or a small FOV is used, probably because part of the galaxies are cut out from the images.
This implies that when the model is applied to another catalog without reliable galaxy radius measurements, one should consider setting the image FOV slightly larger rather than smaller, to make sure that the whole galaxy is contained in the image.

\begin{figure*}
    \epsscale{1.08}
    \plotone{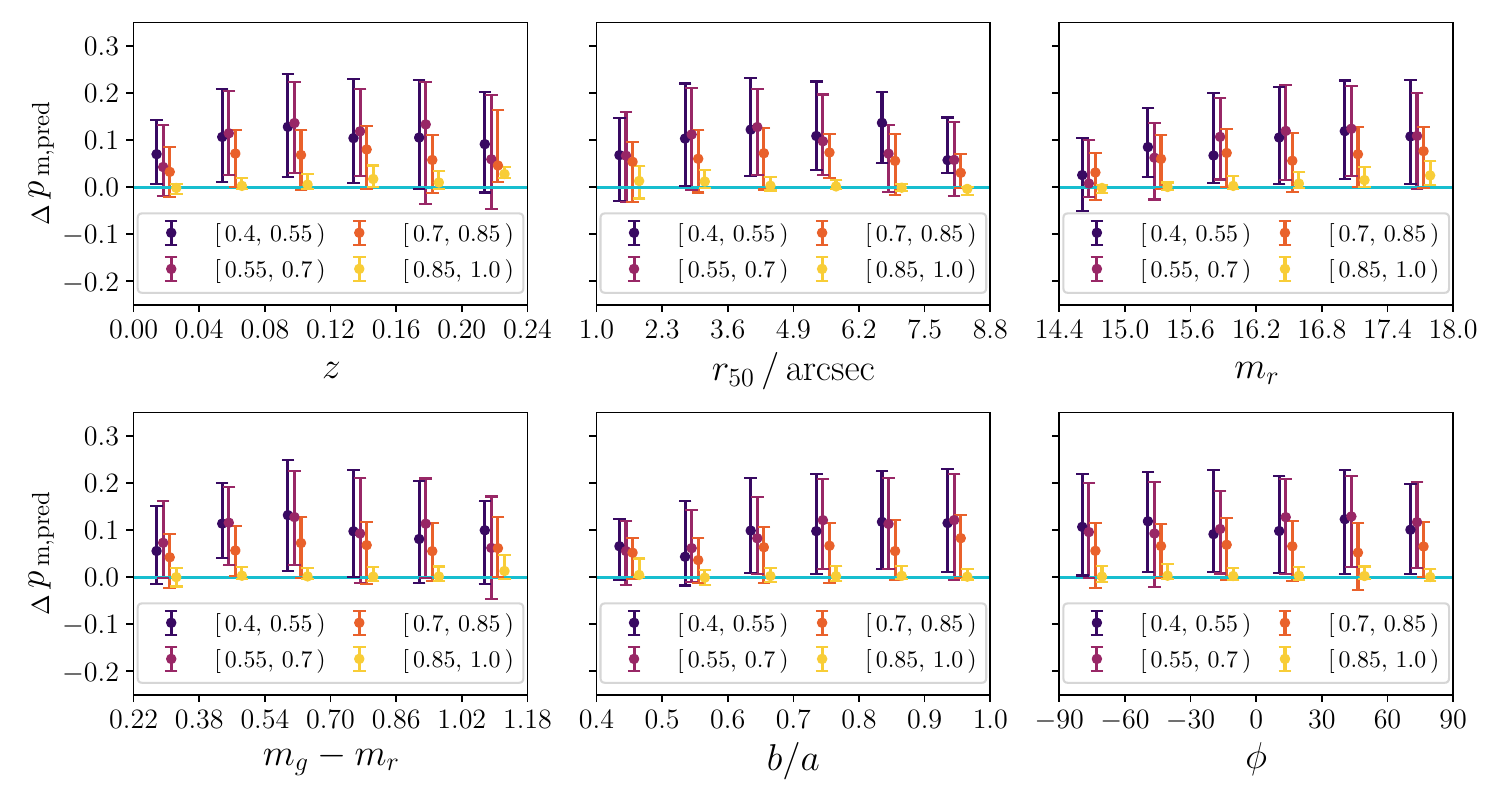}
    \caption{Similar to Figure \ref{fig:sys}, but for the difference in $p_{\rm m}$ when DESI images are used instead of SDSS images.
    The predicted $p_{\rm m}$ increases for almost all types of galaxies.}
    \label{fig:desi-sdss}
\end{figure*}

In Figure \ref{fig:heat}, we show the 2-dim histograms of the true vs predicted GZ1 vote probabilities for the Z-wise and S-wise spirals, which is roughly diagonal with a small dispersion of $\lesssim 0.1$.
Is this a good result?
Since our training data is from the volunteer vote fractions, the dispersion cannot be smaller than the Poisson noise in the data.
Assuming that the votes follow an i.i.d. binomial distribution, the empirical vote fraction will have a standard error of $\Delta p=\sqrt{p(1-p)/n}$, which equals to 0.073 if $p=0.8$ and $n=30$.
Therefore, the performance of our model is already close to the best allowed by the training data.
We also note an excess of galaxies in the $0<p_{\rm z/s,true}<0.1$, $0.1<p_{\rm z/s,pred}<0.3$ bins, due to the relative large number of galaxies with $0<p_{\rm z/s,true}<0.1$.
However, such small $p_{\rm z/s}$ galaxies are rarely relevant to practical applications, since their spin directions are mostly undetermined.

We check whether the classification accuracy systematically depends certain galaxy parameters in Figure \ref{fig:sys}.
All galaxies in the test dataset are binned according to their redshift $z$, half light radius $r_{50}$, $r$ band magnitude $m_r$, color $m_g-m_r$, aspect ratio $b/a$,  orientation $\phi$, as well as the true $p_{\rm m} \equiv \max(p_{\rm z},\,p_{\rm s})$.
We plot the error of $p_{\rm m}$ prediction and the number of galaxies for each bin.
Generally, the error is the smallest for $p_{\rm m}\in [0.85,\,1.0]$, since the Poisson noise follows $\Delta p=\sqrt{p(1-p)/n}$ and decreases with increasing $p$ when $p>0.5$.
We find that our model is overall unbiased, except for the galaxies with large $z$, small $r_{50}$, large $m_r$ and small $b/a$, whose $p_{\rm m}$ is slightly underestimated.
This is unlikely an issue for cosmological applications, however, as the number of such galaxies is relatively small.
Actually, such deviation will not complicate the cosmological analysis, since effectively our model just has a different \textit{selection function} than the humans in GZ1, meaning that a slightly different population of galaxies have their spin directions determined by the classifier.
In principle, one should take this into account to avoid making overconfident claims about our Universe, but the treatment of such selection effects should be similar, regardless of whether the galaxies are selected by humans or machines.

\begin{figure*}
    \epsscale{1.08}
    \plotone{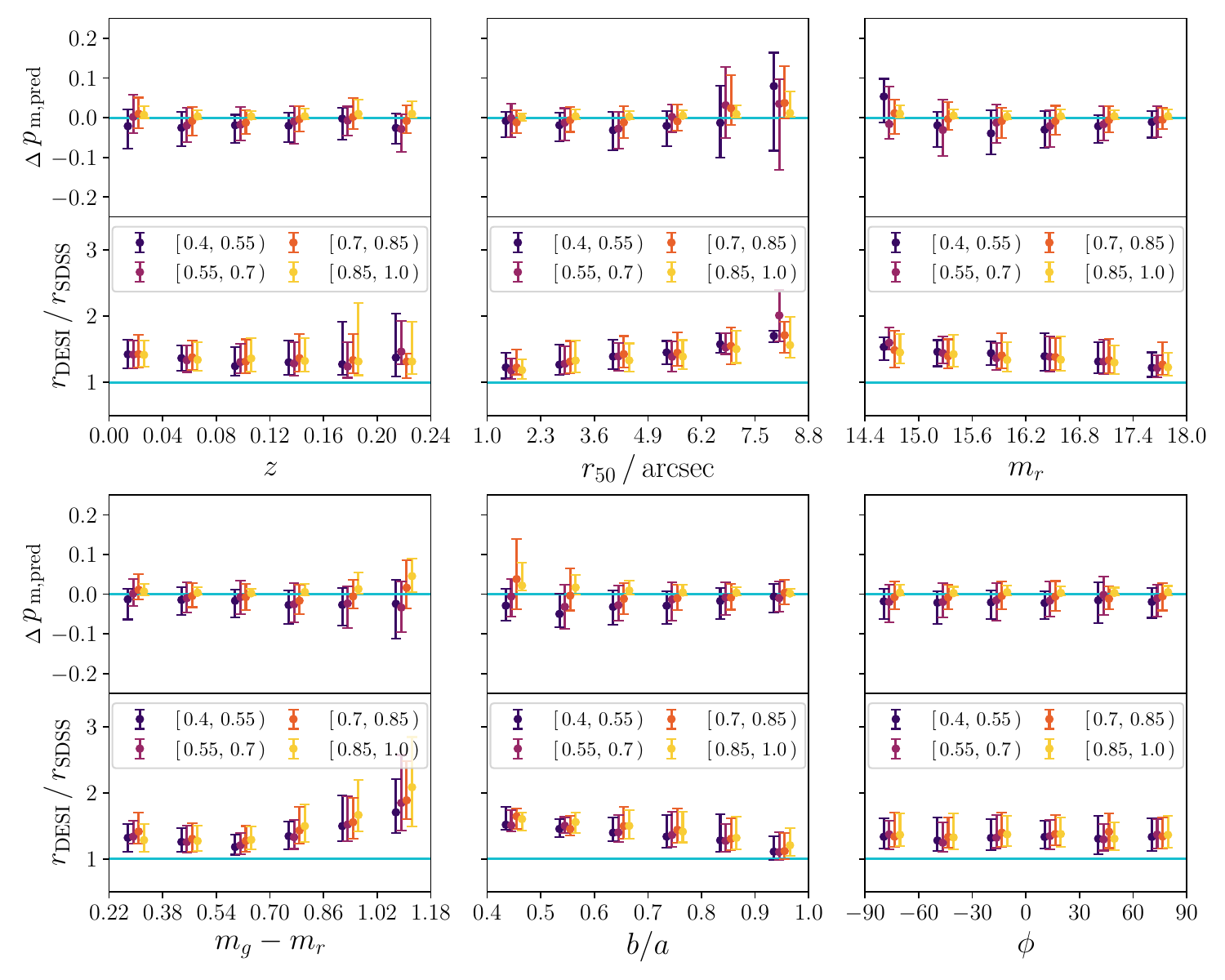}
    \caption{Similar to Figure \ref{fig:sys}, but comparing $p_{\rm m,pred}$ and the measured half light radius, for all the galaxies in the Reduced GZ1 catalog that have one within-1" match in the Preliminary DESI catalog.
    The predicted $p_{\rm m}$ is close for most galaxies despite the difference in the radius measurement, except for the large, red galaxies whose radius measurement differs by about $\gtrsim 100\%$.}
    \label{fig:desi-spec}
\end{figure*}

\section{Application to DESI Images}

\label{sec:desi-image}
\label{sec:pre-desi}

Having trained our CE-ResNet model with SDSS images, we use the model to classify the same galaxies but with DESI images, which generally look redder than SDSS images as the \textit{z} band is used instead of the \textit{i} band for the red channel.
It turns out that our model performs well on these DESI images: we find 31.8\% more spiral galaxies (with $p_{\rm cut}=0.7$) relative to the original human classification (Table \ref{table:pzps}), as the predicted $p_{\rm m}$ increases for almost all kinds of galaxies except those already clearly classified with SDSS images (Figure \ref{fig:desi-sdss}).
Comparing the sample images in Figures \ref{fig:sam-cnn-sdss} and \ref{fig:sam-cnn-desi}, the DESI images are obviously clearer than the SDSS images, enabling better classification of the galaxy morphology.

We also apply our model to the ``Preliminary DESI" catalog, which indeed has a large overlap with the GZ1 catalog, as SDSS contributes most of the current galaxy spectra data before DESI spectra become available.
We find that 150,283 of the 173,097 galaxies in the Reduced GZ1 catalog can be matched to one galaxy within 1" in the Preliminary DESI catalog.
However, these galaxies have slightly different location and radius measurements in DESI Legacy Surveys, which are required to determine the image cut for our model input.
We thus use the ``Preliminary DESI" catalog to validate the accuracy of our model with DESI photometric measurements, since the future DESI spectra catalog will have the galaxy location and radius measured with the same pipeline.

We compare the predicted $p_{\rm m}$ for the galaxies matched between the Reduced GZ1 and Preliminary DESI catalogs.
In principle, $p_{\rm m,pred}$ should be close between these two cases, since we are classifying the same galaxies using the same imaging survey but just with different image cuts.
As shown in Table \ref{table:pzps}, the total number of Z-Spirals and S-Spirals are similar to the Reduced GZ1 catalog with DESI images, with slightly fewer galaxies if one chooses $p_{\rm cut}=0.9$ but more galaxies if one chooses $p_{\rm cut}\in[\,0.5,0.7\,]$.
We find that $\Delta p_{\rm m,pred}$ has a strong correlation with the difference in the measured galaxy radius: most of the galaxies have their radii measured larger by about 30\% in the DESI catalog, which however only leads to a small impact on the predicted $p_{\rm m}$ ($\lesssim 0.05$) since our model is insensitive to a reasonable amount of changes in the image FOV (Figure \ref{fig:stable}).
On the other hand, some large, dim galaxies have $\gtrsim 100\%$ difference in radius measurements, which leads to a larger dispersion in $\Delta p_{\rm m,pred}$.
This should only have limited effects on cosmological applications of our model, however, as the number of such galaxies is relatively small according to Figure \ref{fig:sys}.
Although in practice, one may further improve the performance by applying an empirical correction for the galaxy radii to compensate for the difference between surveys, based on e.g. the general trend of $r_{\rm DESI}/r_{\rm SDSS}$ in Figure \ref{fig:desi-spec}, which we leave for future research.

\section{Chirality Violation Due to Human Bias}

\label{sec:violation}

There is a well-known bias towards S-wise spirals in the GZ1 classification catalog, which has been intensively investigated in a bias study by Galaxy Zoo \citep{lintott2008galaxy,lintott2011galaxy}.
The volunteers were presented flipped galaxy images but still found an excess of S-wise spirals.
It is shown that such bias is due to human selection effect, while the chirality of spiral galaxies is not violated once the human bias is appropriately accounted for \citep{land2008galaxy,hayes2017nature}.

To validate that our CE-ResNet model is parity even, we check the symmetry between the numbers of Z-Spirals and S-Spirals using the classification catalogs in Table \ref{table:pzps}.
We use the following statistics to determine the significance of chirality violation,
\begin{equation}
    T = \frac{n_z-n_s}{\sqrt{n_z+n_s}},
    \label{EQ:TEST}
\end{equation}
which should asymptotically follow the standard Gaussian distribution under the null hypothesis of no chirality violation.
See Appendix \ref{sec:test} for the derivation.

Note that our Reduced GZ1 catalog contains all the $p_{\rm m}>0.3$ galaxies in the full GZ1 catalog, while the $p_{\rm m}\leq0.3$ galaxies are downsampled.
Similar to \citet{hayes2017nature}, we find a $\sim 7\sigma$ asymmetry in the GZ1 Humans classification catalog, which disappears if the same SDSS images are classified by the parity-even CE-ResNet, implying that the asymmetry is due to a slight underestimate of $p_{\rm z}$ relative to $p_{\rm s}$ by GZ1 Humans.
However, when DESI instead of SDSS images are used for the Reduced GZ1 galaxies, there are again slightly more S-wise than Z-wise galaxies with $p_{\rm cut}=0.5$.
This is likely because some of the $p_{\rm z}\sim 0.5$ (with DESI images) galaxies are cut out from the Reduced GZ1 catalog since they may have $p_{\rm z}<0.3$ by GZ1 Humans.
When applied to the Preliminary DESI catalog, our CE-ResNet finds equal number ($<1.8\sigma$ asymmetry) of Z-Spirals and S-Spirals for all different choices of $p_{\rm cut}$, confirming that no real chirality violation between these two types of spirals exists in Nature.

\section{Discussions}

\label{sec:discuss}

In this paper, we present Chirality Equivariant Residual Network (CE-ResNet), a machine learning based model for the classification of Z-wise vs S-wise spirals.
Trained with Galaxy Zoo 1 (GZ1) data, our model gives similar predictions on the chirality of galaxies as the volunteers in the GZ1 project, but can be efficiently applied to the millions or even billions of galaxies that will be mapped in the near feature.
Our model is manifestly parity-even, since basically the same estimator is used to predict the probabilities of Z-Spirals and S-Spirals, using the trick that one gets a S-Spiral if one flips the image of a Z-Spiral.
Heuristically, humans have different \textit{selection functions} for Z- and S-Spirals, while our model learns the average of these two selection functions from humans and use it to classify both types of spirals, which explains how we can train an unbiased classifier from a biased dataset.
We validate our model in Section \ref{sec:train}, and verify that our model can be directly applied to DESI images even though it is trained with SDSS images in Section \ref{sec:desi-image}.
We confirm that no real asymmetry between the numbers of Z-wise and S-wise spirals exists in Section \ref{sec:violation}.

We note a few related works in the literature.
\citet{hayes2017nature} develops an unbiased selector to demonstrate that the excess of S-Spirals in the GZ1 catalog is due to human selection effect.
However, their Unbiased Machine spirality selector finds significantly fewer spirals than GZ1 Humans, such that it cannot maximally extract the information in the survey data for cosmological analysis. 
Recently, \citet{tadaki2020spin} studies the classification of spiral galaxies with CNN.
However, their dataset only includes galaxies that are \textit{unambiguously} Z-spirals, S-spirals and non-spirals, whereas real world survey catalogs also include galaxies with \textit{unclear} morphological type due to the limitation of image quality.
Their model predictions on these unclear galaxies can be undefined as the network has never seen such galaxies during training, making it risky to be directly applied to full survey catalogs.
Also, the classifier in \citet{tadaki2020spin} is not manifestly parity-even, so one should be cautious about the possible inherent asymmetry between Z-Spirals and S-Spirals in their model.

The network used in this paper is equivariant under a reflection of the input image, which eliminates the potential bias caused by the difference between the Z- and S-type estimators.
Additionally, we augment the dataset by a random translation, rotation, scaling and permutation of color channels, to mitigate potential overfitting onto the irrelevant position, orientation, size and color information.
According to Figure \ref{fig:stable}, our network is indeed stable under such transforms.
We note that in principle, one can implement a more advanced equivariant network that manifestly accounts for all the relevant symmetries
\citep[e.g.][]{zhang2019making,weiler2019general,sosnovik2019scale,cesa2022a}.
Also, domain adaption techniques may help to improve the performance when the model needs to be applied to data from different surveys \citep{bendavid2010theory,alexander2021domain}.
We leave these directions for future research, since our current architecture already works well in the various benchmarks demonstrated in this paper.

\section{Acknowledgement}

We thank Jiaxuan Li, Mingwei Ma and Zhehao Xu for helpful discussions.
Ue-Li Pen receives support from Ontario Research Fund-Research Excellence Program (ORF-RE), Natural Sciences and Engineering Research Council of Canada (NSERC) [funding reference number RGPIN-2019-067, CRD 523638-18, 555585-20], Canadian Institute for Advanced Research (CIFAR), Canadian Foundation for Innovation (CFI), the National Science Foundation of China (Grants No. 11929301), Thoth Technology Inc, Alexander von Humboldt Foundation, and the National Science and Technology Council (NSTC) of Taiwan (111-2123-M-001-008-, and 111-2811-M-001-040-).

Computations were performed on the Mist supercomputer at the SciNet HPC Consortium and the SOSCIP Consortium's GPU computing platform.
SciNet is funded by: the Canada Foundation for Innovation; the Government of Ontario; Ontario Research Fund - Research Excellence; and the University of Toronto \citep[][]{Loken_2010}.
SOSCIP is funded by the Federal Economic Development Agency of Southern Ontario, the Province of Ontario, IBM Canada Ltd., Ontario Centres of Excellence, Mitacs and 15 Ontario academic member institutions.

The Legacy Surveys consist of three individual and complementary projects: the Dark Energy Camera Legacy Survey (DECaLS; Proposal ID \#2014B-0404; PIs: David Schlegel and Arjun Dey), the Beijing-Arizona Sky Survey (BASS; NOAO Prop. ID \#2015A-0801; PIs: Zhou Xu and Xiaohui Fan), and the Mayall z-band Legacy Survey (MzLS; Prop. ID \#2016A-0453; PI: Arjun Dey). DECaLS, BASS and MzLS together include data obtained, respectively, at the Blanco telescope, Cerro Tololo Inter-American Observatory, NSF's NOIRLab; the Bok telescope, Steward Observatory, University of Arizona; and the Mayall telescope, Kitt Peak National Observatory, NOIRLab. Pipeline processing and analyses of the data were supported by NOIRLab and the Lawrence Berkeley National Laboratory (LBNL). The Legacy Surveys project is honored to be permitted to conduct astronomical research on Iolkam Du'ag (Kitt Peak), a mountain with particular significance to the Tohono O'odham Nation.

NOIRLab is operated by the Association of Universities for Research in Astronomy (AURA) under a cooperative agreement with the National Science Foundation. LBNL is managed by the Regents of the University of California under contract to the U.S. Department of Energy.

This project used data obtained with the Dark Energy Camera (DECam), which was constructed by the Dark Energy Survey (DES) collaboration. Funding for the DES Projects has been provided by the U.S. Department of Energy, the U.S. National Science Foundation, the Ministry of Science and Education of Spain, the Science and Technology Facilities Council of the United Kingdom, the Higher Education Funding Council for England, the National Center for Supercomputing Applications at the University of Illinois at Urbana-Champaign, the Kavli Institute of Cosmological Physics at the University of Chicago, Center for Cosmology and Astro-Particle Physics at the Ohio State University, the Mitchell Institute for Fundamental Physics and Astronomy at Texas A\&M University, Financiadora de Estudos e Projetos, Fundacao Carlos Chagas Filho de Amparo, Financiadora de Estudos e Projetos, Fundacao Carlos Chagas Filho de Amparo a Pesquisa do Estado do Rio de Janeiro, Conselho Nacional de Desenvolvimento Cientifico e Tecnologico and the Ministerio da Ciencia, Tecnologia e Inovacao, the Deutsche Forschungsgemeinschaft and the Collaborating Institutions in the Dark Energy Survey. The Collaborating Institutions are Argonne National Laboratory, the University of California at Santa Cruz, the University of Cambridge, Centro de Investigaciones Energeticas, Medioambientales y Tecnologicas-Madrid, the University of Chicago, University College London, the DES-Brazil Consortium, the University of Edinburgh, the Eidgenossische Technische Hochschule (ETH) Zurich, Fermi National Accelerator Laboratory, the University of Illinois at Urbana-Champaign, the Institut de Ciencies de l'Espai (IEEC/CSIC), the Institut de Fisica d'Altes Energies, Lawrence Berkeley National Laboratory, the Ludwig Maximilians Universitat Munchen and the associated Excellence Cluster Universe, the University of Michigan, NSF's NOIRLab, the University of Nottingham, the Ohio State University, the University of Pennsylvania, the University of Portsmouth, SLAC National Accelerator Laboratory, Stanford University, the University of Sussex, and Texas A\&M University.

BASS is a key project of the Telescope Access Program (TAP), which has been funded by the National Astronomical Observatories of China, the Chinese Academy of Sciences (the Strategic Priority Research Program ``The Emergence of Cosmological Structures" Grant \#XDB09000000), and the Special Fund for Astronomy from the Ministry of Finance. The BASS is also supported by the External Cooperation Program of Chinese Academy of Sciences (Grant \#114A11KYSB20160057), and Chinese National Natural Science Foundation (Grant \#12120101003, \#11433005).

The Legacy Survey team makes use of data products from the Near-Earth Object Wide-field Infrared Survey Explorer (NEOWISE), which is a project of the Jet Propulsion Laboratory/California Institute of Technology. NEOWISE is funded by the National Aeronautics and Space Administration.

The Legacy Surveys imaging of the DESI footprint is supported by the Director, Office of Science, Office of High Energy Physics of the U.S. Department of Energy under Contract No. DE-AC02-05CH1123, by the National Energy Research Scientific Computing Center, a DOE Office of Science User Facility under the same contract; and by the U.S. National Science Foundation, Division of Astronomical Sciences under Contract No. AST-0950945 to NOAO.


\bibliography{sample631}{}
\bibliographystyle{aasjournal}



\appendix

\section{Sample Galaxy Images}
\label{sec:sample}

\begin{figure*}
    \epsscale{1.17}
    \plotone{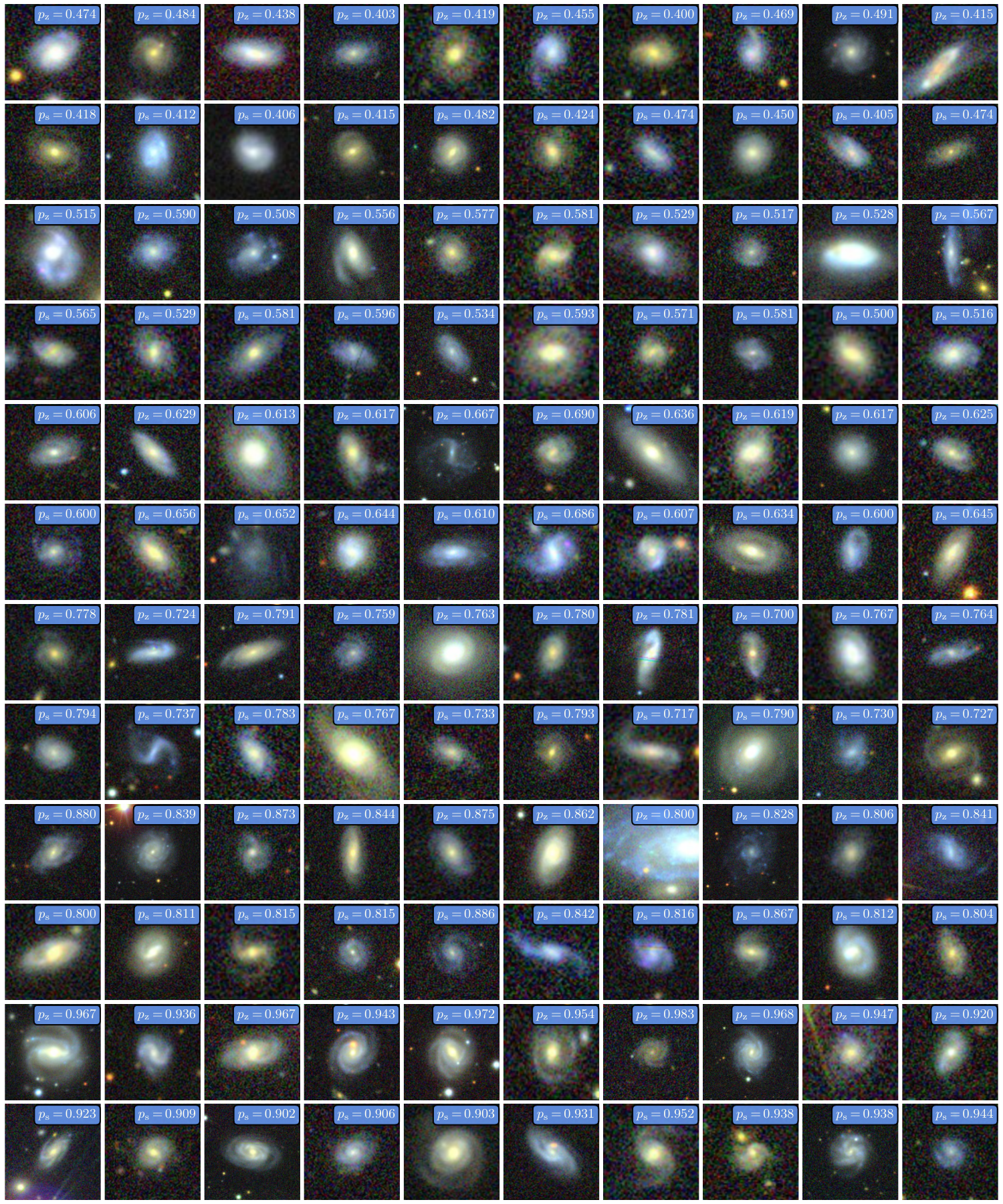}
    \caption{Sample SDSS images for the Reduced GZ1 catalog, with classification probabilities given by GZ1 Humans.}
    \label{fig:sam-gz-sdss}
\end{figure*}

\begin{figure*}
    \epsscale{1.17}
    \plotone{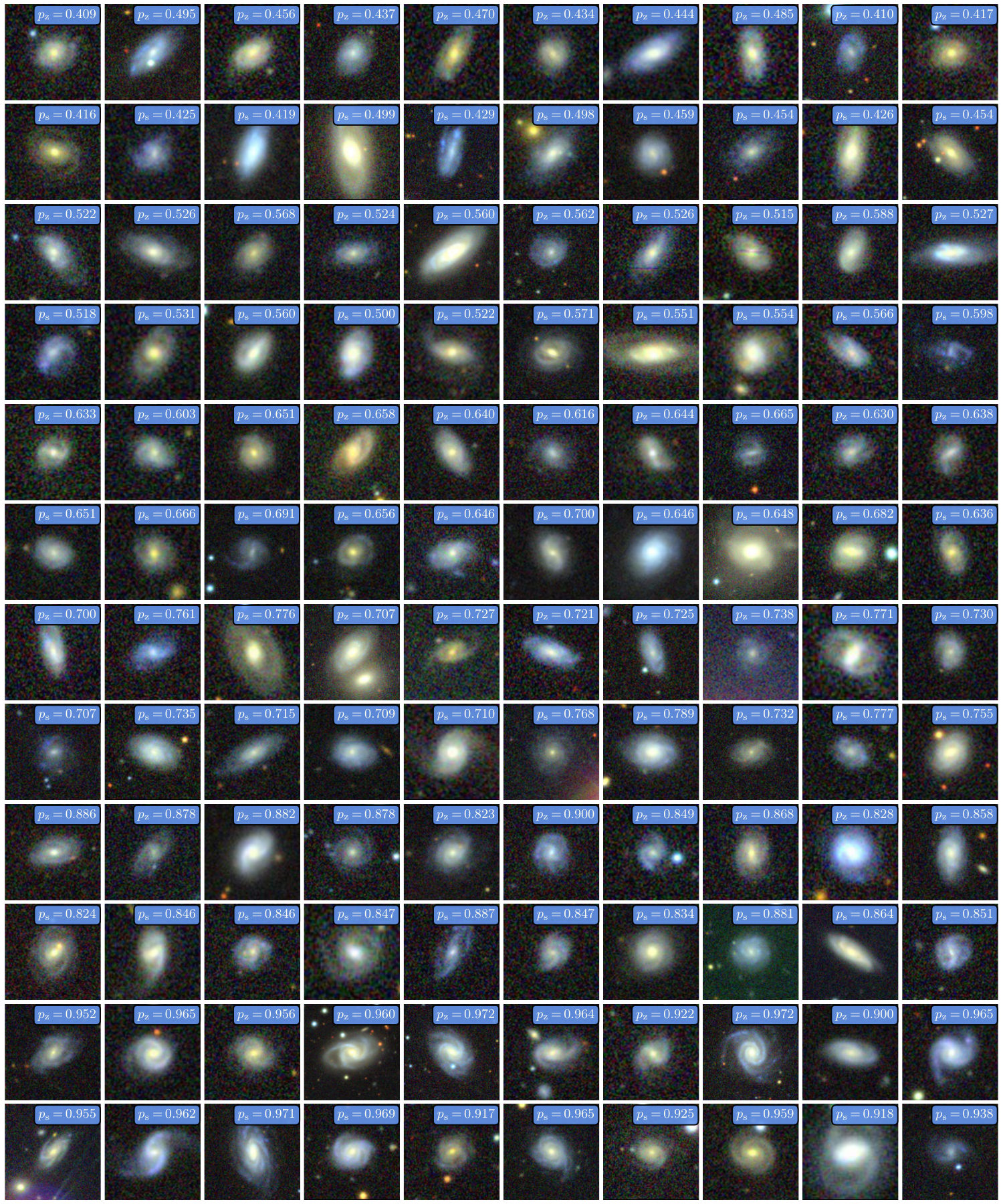}
    \caption{Sample SDSS images for the Reduced GZ1 catalog, with classification probabilities predicted by CE-ResNet.}
    \label{fig:sam-cnn-sdss}
\end{figure*}

\begin{figure*}
    \epsscale{1.17}
    \plotone{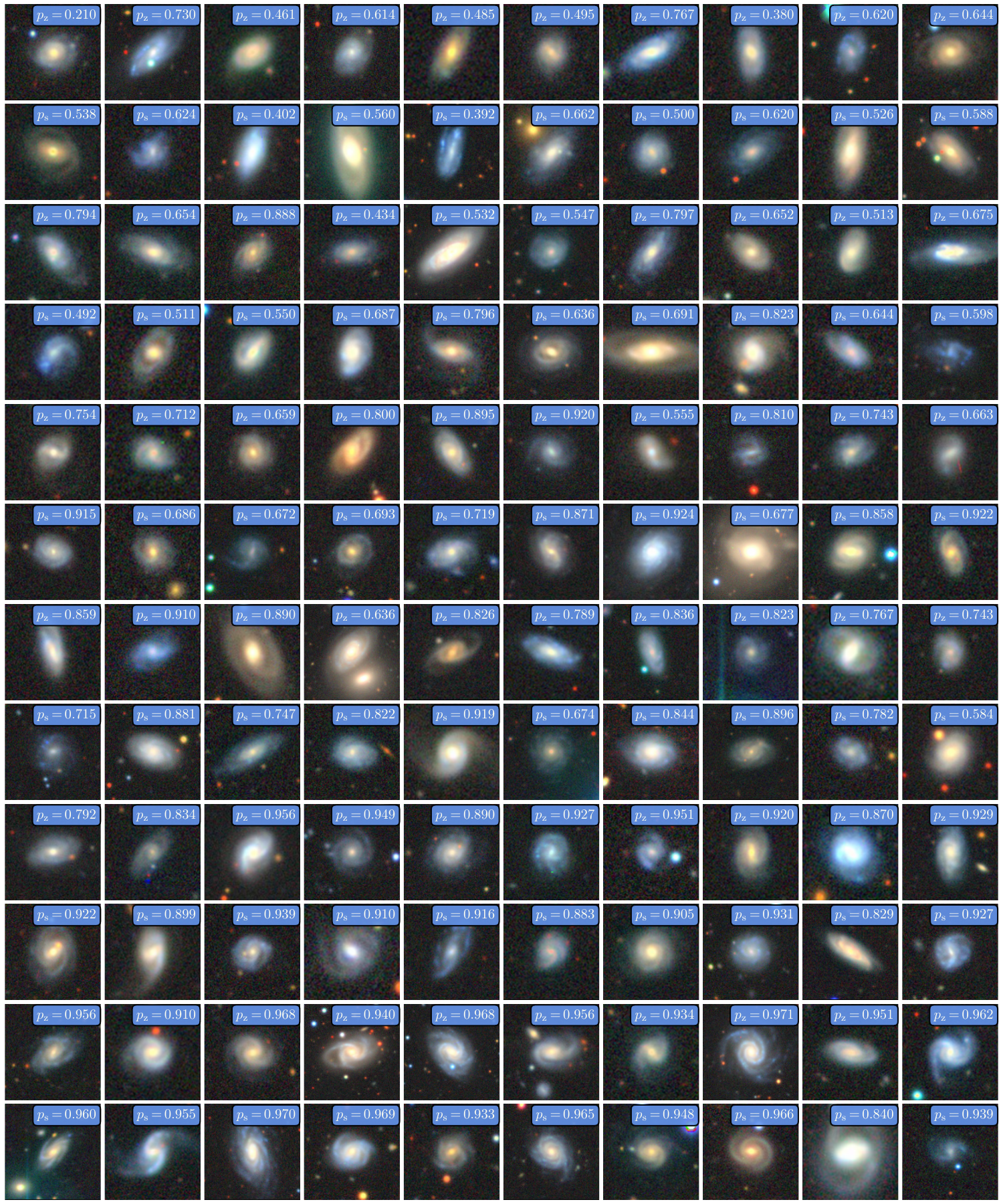}
    \caption{The same galaxies as Figure \ref{fig:sam-cnn-sdss} but classified with DESI images.
    The better imaging quality enables more confident morphology classification for many galaxies.}
    \label{fig:sam-cnn-desi}
\end{figure*}

We show sample galaxy images for the various classification catalogs in Figures \ref{fig:sam-gz-sdss}-\ref{fig:sam-cnn-desi}.
They all come from the GZ1 catalog, but are classified by GZ1 Humans with SDSS images, by CE-ResNet with SDSS images, and by CE-ResNEt with DESI images, respectively.
In Figures \ref{fig:sam-gz-sdss}-\ref{fig:sam-cnn-sdss}, we randomly choose 10 galaxies in each $p_{\rm z/s}$ bin of width 0.1.
Figure \ref{fig:sam-cnn-desi} shows the same galaxies as Figure \ref{fig:sam-cnn-sdss}; it is clear that the better images of DESI lead to the increase of $p_{\rm z/s}$ prediction.

\section{Derivation of Equation \ref{EQ:TEST}}

\label{sec:test}

Suppose that we have observed $n$ galaxies in total, of which $n_z$ are Z-wise spirals and $n_s$ are S-wise spirals.
Let the operator $\mathbb{1}_z$ equal to 1 if a galaxy is Z-wise and 0 otherwise, and similarly for the operator $\mathbb{1}_s$.
Under the null hypothesis of no chirality violation, we have ${\rm Exp}\,(\mathbb{1}_z-\mathbb{1}_s)=0$, and
\begin{equation}
    {\rm Var}\,(\mathbb{1}_z-\mathbb{1}_s)={\rm Exp}\,[(\mathbb{1}_z-\mathbb{1}_s)^2]-[{\rm Exp}\,[(\mathbb{1}_z-\mathbb{1}_s)]^2=\left[ \frac{n_z+n_s}{n} \right]_{n\to\infty},
\end{equation}
where we have used ${\rm Exp}\,(\mathbb{1}_z\mathbb{1}_s)=0$ as one galaxy cannot be both Z-wise and S-wise.
According to the central limit theorem, in the limit of large $n$,
\begin{equation}
    \sqrt{n} \left<\mathbb{1}_z-\mathbb{1}_s\right>_n \sim N\left[0,\,\left(\sqrt{\frac{n_z+n_s}{n}}\right)^2\right],
\end{equation}
from which it is straightforward to show that the $T$ statistics in Equation \ref{EQ:TEST} follows the standard Gaussian distribution.

\end{document}